\documentclass[%
 reprint,
 superscriptaddress,
 longbibliography,
 amsmath,amssymb,
 aps,
 prl,
]{revtex4-2}

\usepackage{tabularx}
\usepackage{booktabs}
\usepackage{upgreek}
\usepackage{nicefrac}
\usepackage{hyperref}
\usepackage{natbib}
\usepackage{xcolor}
\usepackage{graphicx}% Include figure files
\usepackage{dcolumn}% Align table columns on decimal point
\usepackage{bm}% bold math

\usepackage{multirow}

\usepackage{float}

\usepackage{lineno}

\begin{document}

\preprint{APS/123-QED}

\title{Charge Radius of Neutron-deficient $^{54}$Ni and Symmetry Energy Constraints Using the Difference in Mirror Pair Charge Radii}% Force line breaks with \\
\author{Skyy V. Pineda}
 \email{pineda@frib.msu.edu}
\affiliation{National Superconducting Cyclotron Laboratory, Michigan State University, East Lansing, Michigan 48824, USA}
\affiliation{Department of Chemistry, Michigan State University, East Lansing, Michigan 48824, USA}

\author{Kristian K\"onig}
\affiliation{National Superconducting Cyclotron Laboratory, Michigan State University, East Lansing, Michigan 48824, USA}

\author {Dominic M. Rossi}
\affiliation{Institut f\"ur Kernphysik, Technische Universit\"at Darmstadt, 64289 Darmstadt, Germany}
\affiliation{GSI Helmholtzzentrum f\"ur Schwerionenforschung mbH, Planckstr. 1, 64291 Darmstadt, Germany}

\author{B. Alex Brown}
\affiliation{National Superconducting Cyclotron Laboratory, Michigan State University, East Lansing, Michigan 48824, USA}
\affiliation{Department of Physics and Astronomy, Michigan State University, East Lansing, Michigan 48824, USA}

\author {Anthony Incorvati}
\affiliation{National Superconducting Cyclotron Laboratory, Michigan State University, East Lansing, Michigan 48824, USA}
\affiliation{Department of Physics and Astronomy, Michigan State University, East Lansing, Michigan 48824, USA}

\author{Jeremy Lantis}
\affiliation{National Superconducting Cyclotron Laboratory, Michigan State University, East Lansing, Michigan 48824, USA}
\affiliation{Department of Chemistry, Michigan State University, East Lansing, Michigan 48824, USA}

\author{Kei Minamisono}
\email{minamiso@nscl.msu.edu}
\affiliation{National Superconducting Cyclotron Laboratory, Michigan State University, East Lansing, Michigan 48824, USA}
\affiliation{Department of Physics and Astronomy, Michigan State University, East Lansing, Michigan 48824, USA}

\author {Wilfried N\"ortersh\"auser}
\affiliation{Institut f\"ur Kernphysik, Technische Universit\"at Darmstadt, 64289 Darmstadt, Germany}

\author{Jorge Piekarewicz}
\affiliation{Department of Physics, Florida State University, Tallahassee, Florida 32306, USA}

\author{Robert Powel}
\affiliation{National Superconducting Cyclotron Laboratory, Michigan State University, East Lansing, Michigan 48824, USA}
\affiliation{Department of Physics and Astronomy, Michigan State University, East Lansing, Michigan 48824, USA}

\author {Felix Sommer}
\affiliation{Institut f\"ur Kernphysik, Technische Universit\"at Darmstadt, 64289 Darmstadt, Germany}

\date{\today}% It is always \today, today,

\begin{abstract}
The nuclear root-mean-square charge radius of $^{54}$Ni was determined with collinear laser spectroscopy to be $R(^{54}$Ni) = 3.737\,(3)~fm. In conjunction with the known radius of the mirror nucleus $^{54}$Fe, the difference of the charge radii was extracted as $\Delta R_{\rm ch}$ = 0.049\,(4)~fm. Based on the correlation between $\Delta R_{\rm ch}$ and the slope of the symmetry energy at nuclear saturation density ($L$), we deduced $20 \le  L \le 70$\,MeV. The present result is consistent with the $L$ from the binary neutron star merger GW170817, favoring a soft neutron matter EOS, and barely consistent with the PREX-2 result within 1$\sigma$ error bands. Our result indicates the neutron-skin thickness of  $^{48}$Ca as 0.15\,-\,0.19\,fm. %Finally, we explore the impact of our result on the neutron skin thickness of $^{48}$Ca, which will soon be announced by the CREX collaboration.
\end{abstract}

\maketitle

%\section{Introduction}
%\label{sec:Introduction}

\paragraph{Introduction ---}
\label{sec:Introduction}
Knowledge of the slope  of the symmetry energy $L$ in the nuclear equation of state (EOS) is critical for the extrapolation to the higher densities \cite{Brown.2000} that are required to predict the properties of both super-heavy nuclei and neutron stars \cite{hor01, Steiner.2013, ste13}. In the case of neutron stars, the ``softness" or ``stiffness" of the EOS has a direct link to the neutron star radius\,\cite{Lattimer:2006xb}. Note that a stiff EOS indicates that the pressure increases rapidly with increasing density. Conceptually, the symmetry energy is closely related to the difference between the energy per nucleon of pure neutron matter and symmetric nuclear matter. Given that symmetric nuclear matter saturates, $L$ is proportional to the pressure of pure neutron matter at nuclear saturation density $\rho_0$ \cite{Brown.2017}. Different parameterizations of Skyrme energy density functionals show dramatic variations in the stiffness of the EOS\,\cite{Brown.2000}, therefore making the extrapolations to higher densities uncertain. The stiffness of the EOS in the vicinity of $\rho_0$ is controlled by $L$, and although $L$ cannot be directly determined through experiment, the neutron skin thickness $\Delta R_{\rm np}$ of neutron rich nuclei is strongly correlated to $L$ \cite{RocaMaza:2011pm,Reinhard.2016}, which may then be used to set boundaries on its value \cite{Brown.2017}. 

The lead radius experiments PREX-1 \cite{Abrahamyan.2012} and PREX-2 \cite{adh21} provide a direct probe of neutron densities via parity violating electron scattering. Given that the weak charge of the neutron is much larger than that of the proton, it paves an electroweak avenue to constrain the density dependence of the symmetry energy. Other electromagnetic methods involve a correlation between the electric dipole polarizability and the $\Delta R_{\rm np}$\,\cite{Reinhard.2010, Piekarewicz.2011}. Such measurements have been performed in $^{208}$Pb\,\cite{tam11, Roca-Maza.2013}, $^{48}$Ca\,\cite{Birkhan:2016qkr}, and in radioactive $^{68}$Ni \cite{Rossi.2013}. Besides terrestrial experiments, the binary neutron star merger GW170817 has placed important constraints on the EOS through the analysis of the tidal polarizability (or deformability)\,\cite{Abbott.2018}. %The gravitational wave form, the outcome of the merger (e.g., prompt or delay collapse into a black hole), and the post-merger kilonova emission, are among other effects sensitive to the EOS. 
Various studies have aimed to translate the measurements on the neutron star merger into constraints on the EOS of dense neutron matter. However, whether the EOS is soft or stiff---which in turn translates into smaller or larger neutron star radii, respectively---is still under debate \cite{Abbott.2017, Abbott.2018, Fattoyev:2017jql, Tews:2018chv, Tsang.2019, Zhang.2020, Shen.2020, Hu.2020, Li.2021, est21}.

Another purely electromagnetic method to constrain $L$ has been introduced in \cite{Brown.2017}, where the $\Delta R_{\rm np}$ is deduced from the difference in charge radii between a mirror pair. Assuming perfect charge symmetry, the neutron radius of a given nucleus should be equal to the proton 
radius of the corresponding mirror nucleus. The $\Delta R_{\rm np}$ can then be obtained from the difference $\Delta R_{\rm ch}$ of the
root-mean-square (rms) charge radii $R_{\rm ch}$ of mirror nuclei \cite{Brown.2017, Yang.2018} as $\Delta R_{\rm np} = R_{\rm ch}\left(^A_ZX_N \right) - R_{\rm ch}\left(^A_NY_Z \right) = \Delta R_{\rm ch}$, where $A\!=\!N+Z$ is the mass number, and $N$ and $Z$ are the neutron and proton number, respectively. In reality, however, the charge symmetry is broken by the Coulomb interaction that pushes protons out relative to neutrons, leading to a weaker correlation between $\Delta R_{\rm np}$ and $\Delta R_{\rm ch}$. It turns out that $\Delta R_{\rm ch}$ is strongly correlated with $|N - Z|$ $\times$ $L$ even when $|N - Z|$ is small. On the other hand, $\Delta R_{\rm np}$ depends on both $|N - Z| \times L$ and the symmetry energy with the $L$ dependence dominating at large $|N - Z|$ \cite{Brown.2017}. Such experiments provide a clean and largely model independent complement to the parity violating asymmetry experiments. In the present study, the mirror charge radii formalism is applied to the $^{54}\mathrm{Ni}$-$^{54}\mathrm{Fe}$ pair. The rms charge radius of $^{54}$Ni was determined for the first time and then combined with the known radius of stable $^{54}$Fe\,\cite{Fricke.2004}. Although this pair has a smaller $|N - Z|$ = 2 relative to our previous measurement on the $^{36}\mathrm{Ca}$-$^{36}\mathrm{S}$ mirror pair \cite{Brown.2020}, the precise determination of the charge radius of $^{54}$Ni provides a meaningful constraint on $L$, with input from modern nuclear models. 

\paragraph{Experiment ---}
\label{sec:Experimental}
This experiment took place at the National Superconducting Cyclotron Laboratory at Michigan State University. A $^{58}$Ni primary beam was impinged upon a beryllium target and the produced $^{54}$Ni($I^\pi=0^+$, $T_{1/2}=114$ ms) beam was filtered out using the A1900 fragment separator. The isolated $^{54}$Ni beam was then thermalized in a gas cell \cite{Sumithrarachchi.2020}, extracted at an energy of 30 keV and transported to the BECOLA facility \cite{Minamisono.2013, Rossi.2014}. A typical rate of Ni$^{+}$ ions at the entrance of the BECOLA was typically 400/s. At BECOLA the Ni beam was captured, cooled and bunched in a radio frequency quadrupole (RFQ) ion trap \cite{bar17}.
%to improve the beam emittance and to perform time-resolved fluorescence measurements on the bunched beam \cite{Rossi.2014} in a collinear geometry. 
The ion beam was extracted from the RFQ at an approximate energy of 29850 eV.
Then the beam was neutralized in-flight in a charge-exchange cell (CEC) \cite{Klose.2012}. The typical neutralization efficiency was 50\%, and the metastable $3d^94s$ $^3\mathrm{D}_3$ state was populated, which was estimated by a simulation to be 15\% \cite{Ryder.2015} of the total population. A small scanning potential (typically 50 V) was applied to the CEC to change the velocity of the incident ion beam and thus of the atom beam. This in turn Doppler-shifted the laser frequency in the rest frame of the atoms, and effectively scanned the laser frequency to measure the hyperfine spectrum. Ions in the metastable state were excited with 352-nm laser light to the $3d^94p$ $^3\mathrm{P}_2$ state, and fluorescence light was recorded as a function of the scanning voltage with a mirror-based fluorescence detection system\cite{Minamisono.2013, mas20}. A background suppression factor of 2$\times$10$^5$ was achieved by performing time-resolved fluorescence measurements with the bunched beam \cite{Rossi.2014,cam02, Nieminen.2002}.

%A time-resolved fluorescence measurement with the bunched beam \cite{Rossi.2014} was performed leading to a background suppression factor of 2$\times$10$^5$ \cite{cam02, Nieminen.2002}. A Sirah Matisse TS Ti:sapphire ring laser was used to produce 704-nm light. A Spectra Physics wave train generated 352-nm light by frequency doubling the 704-nm light. 

\begin{figure}[t]
\centering
    \includegraphics[width = 0.5\textwidth]{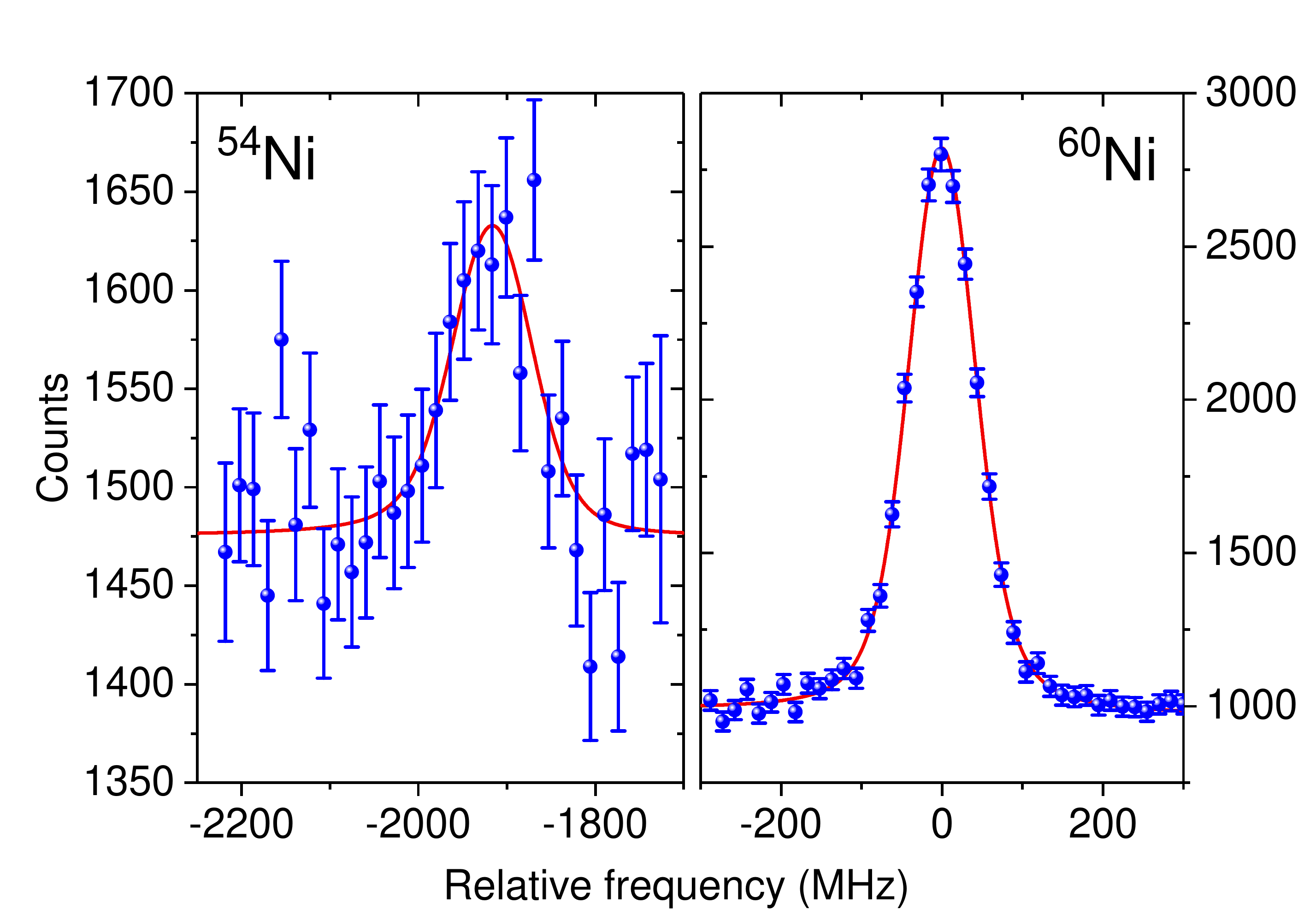}
    \caption{Resonance spectra for $^{54}$Ni (left) and $^{60}$Ni (right) relative to the rest-frame transition frequency of $^{60}$Ni. The solid line is the fit to the data.}
    \label{fig:Res_spectra}
\end{figure}
A Penning Ionization Gauge (PIG) ion source \cite{Ryder.2015} was used to generate beams of stable $^{58, 60}$Ni isotopes, and spectroscopy was performed every 4-6 hours throughout the data taking time for $^{54}$Ni. The resonance frequencies of $^{58, 60}$Ni were used as the reference for the extraction of the $^{54}$Ni isotope shift as well as to determine the kinetic beam energy with $10^{-5}$ relative accuracy \cite{Koenig.2021}. When changing between the isotopes, the laser frequency was adjusted to perform spectroscopy at the same beam energy. The applied laser frequencies were referenced against molecular iodine transition lines \cite{Powel.2021}.

\paragraph{Experimental Results ---}
\label{sec:expresult}
The observed resonance line of $^{54}$Ni is shown in Fig.\,\ref{fig:Res_spectra}~(left). A Voigt function with an exponential low-energy tail to describe the asymmetry caused by inelastic collisions with the sodium vapor \cite{Klose.2012} was used to fit the $^{54}$Ni spectrum, and the fit result is shown as a solid line. The asymmetry parameter and the Lorentz width of the Voigt function were fixed to those obtained from the reference measurements on $^{58}$Ni and $^{60}$Ni. A typical spectrum of $^{60}$Ni is shown in Fig.\,\ref{fig:Res_spectra}~(right) as an example of a stable isotope measurement. 

The isotope shifts defined as $\delta\nu^{A,A^\prime} = \nu^{A} - \nu^{A^\prime}$ were extracted and summarized in Table \ref{tab:King-plot}. The uncertainty is dominated by the statistical uncertainty of the $^{54}$Ni resonance centroid (7.5 MHz). A discussion of the systematic uncertainty contributions is detailed in \cite{Koenig_Kingfit.2021}. From the obtained isotope shifts, the differential mean square (ms) charge radius was extracted as $\delta \langle r^2 \rangle^{A,A'} = (\delta \nu^{A,A'} - \mu^{A,A'} K_\alpha)/F + \mu^{A,A'} \alpha$ \cite{Hammen.2018} with the offset parameter $\alpha$, the field-shift factor $F$, the offset-dependent mass-shift factor $K_{\alpha}$, and $\mu^{A,A'}=(m_{A}-m_{A'})/\{(m_{A}+m_e) (m_{A'}+m_e)\}$, where $m_{A}$ and $m_{A'}$ are the nuclear masses, and $m_e$ is the electron mass. The $F$ and $K_\alpha$ were separately determined \cite{Koenig_Kingfit.2021} by the King-fit analysis \cite{King.1984} using re-measured isotope-shifts of the stable isotopes, and are listed in Tab.\,\ref{tab:King-plot} for $^{58}$Ni and $^{60}$Ni as reference isotopes. Here, the offset parameter $\alpha$ was chosen to remove the correlation between the field- and mass-shift parameters in the linear regression. The obtained differential ms and the rms charge radii are also listed in Tab.\,\ref{tab:King-plot}.
\begin{table}[b]
    \centering
    \caption{Isotope shift, atomic parameters, differential ms and rms charge radii of $^{54}$Ni for $A'=58$ and $A'=60$ as the reference isotope are summarized.}
    \begin{tabularx}{0.48\textwidth}{>{\raggedright\arraybackslash}X >{\centering\arraybackslash}X >{\centering\arraybackslash}X}
        \hline
        \multicolumn{1}{c}{}	& \multicolumn{1}{c}{$A'=58$} &  \multicolumn{1}{c}{$A'=60$} \\
        \hline
        $\delta\nu^{54,A^\prime}$\,/\,$\mathrm{MHz}$ & -1410.4\,(8.2) & -1919.7\,(7.9)\\
        \hline
        $\alpha\,/\,\mathrm{u}\,\mathrm{fm}^{2}$ & 417 & 388\\
        $K_\alpha\,/\,\mathrm{GHz/u}$ & 929.8\,(2.2) & 954.0\,(3.5)\\
        $F\,/\, \mathrm{MHz/fm^2}$ &-767\,(70)   & -804\, (66)\\
        \hline
        $\delta\big\langle r^2\big\rangle^{54,A'}$/\,fm$^2$ & -0.235\,(29)  & -0.522\,(20)\\
        $R(^{54}$Ni)  /\,fm  &3.738\,(4)  & 3.737\,(3) \\
        \hline
    \end{tabularx}
    \label{tab:King-plot}
\end{table} 
The differential ms charge radii were used together with the known rms charge radii for reference isotopes to determine the rms charge radius of $^{54}$Ni as $R(^{54}{\rm Ni})= \{(R(^{A^\prime}{\rm Ni}))^2 + \delta \langle r^2\rangle^{54,A'}\}^{1/2}$. 
The rms charge radii of $^{58}$Ni, $^{60}$Ni and $^{54}$Fe were evaluated by combining tabulated values \cite{Fricke.2004} for the Barrett radii $R_{k\alpha}$ from muonic spectroscopy and for the ratio of the radial moments $V_2$ from electron scattering, which yields the model-independent rms charge radii $R_{\rm ch}$~=~$R_{k\alpha}/V_2$ as 3.7698\,(16)\,fm, 3.8059\,(17)\,fm and 3.6880\,(17)\,fm, respectively.
%The rms charge radii of $^{58}$Ni, $^{60}$Ni and $^{54}$Fe were evaluated \cite{Koenig_Kingfit.2021} from the tabulated muonic spectroscopy and electron-scattering data in \cite{Fricke.2004} as 3.7698\,(16) fm, 3.8059\,(17) fm and 3.6879\,(17) fm, respectively. 
%
With the rms charge radii of $^{54}\mathrm{Fe}$ the difference in mirror charge radii was determined to be $\Delta R_{\mathrm{ch}} = R({^{54}{\rm Ni}})-R({^{54}{\rm Fe}}) = 0.049\,(4)\,\mathrm{fm}.$

\paragraph{Theoretical radii ---}
Predictions were made for the difference in charge radii of $^{54}$Ni and $^{54}$Fe using the 48 Skryme energy-density functionals (EDF) \cite{Brown.2017} and the covariant density-functional (CODF) theory where a correlation between $  \Delta R_{\rm ch}  $ and $  L  $ was also observed \cite{Yang.2018}. 

For the $  A = 36  $  mirror pair \cite{Brown.2020}, it was found that the Skyrme results are sensitive to the isoscalar (IS) or the isoscalar plus isovector (IS$+$IV) forms of the spin-orbit potential. However, the present $  A = 54  $ pair turns out to be insensitive to the forms. The IS results is about 0.003\,fm larger in $  \Delta R_{\rm ch}  $, which is negligible, and therefore we adapted the standard IS$+$IV form in this paper. 

The Skyrme \cite{Brown.2017} and CODF \cite{del10} calculations include the relativistic spin-obit (RSO) correction to the charge radius \cite{hor12}, and were performed for spherical nuclei. It is known that the quadrupole correlations increase the rms radii when the saturation condition of isoscalar nuclear matter is taken into account \cite{ber19}. In the present work, the quadrupole deformation effects were taken into account as a correction, which is discussed in the following.  

The Bohr Hamiltonian starts with an expansion of the nuclear surface in terms of of its multipole degrees of freedom
\begin{equation}
R(\theta ,\phi ) = R_{0} \left[ 1 + \displaystyle\sum _{\lambda ,\mu } \alpha _{\lambda ,\mu } 
Y_{\lambda ,\mu }(\theta ,\phi ) \right],     
\label{eq1}
\end{equation}
where $  R_{0}  $ is the radius of the nucleus when it has the spherical equilibrium shape, and $  Y_{\lambda,\mu}  $ is the spherical harmonic. The integrals of Eq.\,(\ref{eq1}) involve $\beta ^{2} = \sum_{\lambda  \geq 2} \sum_{\mu } \mid\alpha _{\lambda,\mu }\mid ^{2}$. To order $\beta^{2}$, the volume integral of Eq.\,(\ref{eq1}) is $I_{0} = \{R_{0}^{3} (4\pi  +3 \alpha _{0} \, \sqrt{4\pi }  + 3 \beta ^{2})\}/3$. Proton ($  q = p  $), neutron ($  q = n  $) and matter ($  q = m  $) distributions are distinguished by using $  R_{0q}  $, $  \alpha _{0q}  $ and  $\beta_{q}$. For the matter density, if we impose the condition of saturation (that the average interior density remains constant), then the volume must be conserved, $  I_{0} = 4\pi R_{0m}^{3}/3  $. This condition can be imposed by having 
\begin{equation}
\alpha _{0m} = -\frac{\beta _{m}^{2}}{\sqrt{4\pi }}.     
\label{eq4}  
\end{equation}
To order $\beta^{2}$, the $  r^{2}  $ integral is $I_{2} = \{R_{0}^{5} (4\pi  + 5 \alpha _{0} \, \sqrt{4\pi }  + 10 \beta ^{2})\}/5$. With the condition of volume conservation from Eq.\,(\ref{eq4}), the matter ms radius is
\begin{equation}
\left<r^{2}\right>_{m} =  \frac{I_{2}}{I_{0}} =  \left<r^{2}\right>_{0m} \left[ 1+\frac{5}{4\pi } \beta _{m}^{2} \right],       
 \label{eq6}
\end{equation}
where $\left<r^{2}\right>_{0m} = 3R_{0m}^{2}/5$ is the ms radius with no deformation.
If $\beta_{p}$ = $\beta_{n}$ = $\beta_{m}$, then we can use Eq.\,(\ref{eq6}) for protons. 
But if $\beta_p \neq \beta_n$, one must make some assumptions about the $\alpha_0$ term.
If we take $\alpha_{0p} = \alpha_{0n} = \alpha_{0m}$ for the volume correction, then
\begin{align}
\left<r^{2}\right>_{p} & = \left<r^{2}\right>_{0p}  \left[ 1 + \frac{2 \alpha _{0p} }{\sqrt{4\pi }} + \frac{7}{4\pi } \beta _{p}^{2} \right] \nonumber\\
&=  \left<r^{2}\right>_{0p}  \left[ 1 - \frac{2}{4\pi } \beta _{m}^{2} + \frac{7}{4\pi } \beta _{p}^{2} \right].       
\label{eq8}
\end{align}

For $\lambda$ = 2, the $\beta_{p}$ are related to the $  B(E2,\uparrow)_{p}  $ for 0$^{ + }$ to 2$^{ + }$ (in units of $e^{2}$) by $\beta _{p} =4\pi \sqrt{B(E2,\uparrow)_{p}}/(5 a_{q}\left<r^{2}\right>_{0p})$, where $  a_{q} = Z  $ for protons. For $\beta_{n}$ and $\beta_{m}$ we have equivalent expressions involving neutrons with $  a_{q} = N  $ and matter with $  a_{q} = A  $. The calculated $  B(E2,\uparrow)_{p}  $ can be compared to electromagnetic experimental results, whereas  $  B(E2,\uparrow)_{n}  $ and $  B(E2,\uparrow)_{m}  $ are much less well known from hadronic scattering experiments. The $  B(E2,\uparrow)_{q}  $ was calculated from the full-basis configuration interaction calculations in the $  fp  $ shell-model space with the GFPX1A Hamiltonian \cite{hon05}. The basic quantities calculated are the model-space matrix elements denoted by $  A_{q}  $. The full matrix element $  M_{q} = \sqrt{B(E2,\uparrow)_{q}}  $ was obtained with ``effective charges" $  \delta _{cv}  $ that arise from 2$\hbar\omega$ admixtures of core nucleons ($  c  $) induced by the valence $  (fp)  $ nucleons ($  v  $) as $M_{p} = A_{p} (1+\delta _{pp}) + A_{n} \delta _{pn}$ and $M_{n} = A_{n} (1+\delta _{nn}) + A_{p} \delta _{np}$. Here $  M_{m} = M_{p} + M_{n}  $. We take the approximation that $  \delta _{pp} = \delta _{nn}  $ and $  \delta _{pn} = \delta _{np}  $. The isoscalar effective charge was evaluated using the data in \cite{hon04} as $  \delta _{0} = \delta _{pp}+\delta _{pn}  \approx 1.00$. Comparing the $  fp  $ model-space calculations to data for mirror transitions in $  A=51  $ \cite{rie04} found $  \delta _{1} \approx -0.60  $. Therefore, $  \delta _{pn} = 0.80  $ and $  \delta _{pp} = 0.20  $ were used. For the $  A_{q}  $ the radial matrix elements are calculated with harmonic-oscillator radial wavefunctions with $  \hbar \omega  = 45 A^{-1/3} - 25 A^{-2/3}  $ \cite{blo68}.

For $^{54}$Fe we obtain $  B(E2,\uparrow)_{p} = 619\,$ e$^{2}$\,fm$^{4}$ compared to the experimental value of 640(13) e$^{2}$\,fm$^{4}$ \cite{yur04}. The  value for $^{54}$Ni is $  B(E2,\uparrow)_{p} = 467\,$ e$^{2}$\,fm$^{4}$ compared to an experimental value of $<$\,800\,e$^{2}$\,fm$^{4}$ \cite{yur04}. There is additional $E2$ strength up to about 4 MeV in excitation energy in the calculations and in experiment \cite{yur04}. Use of the collective model for the radius change can be justified by treating the lowest 2$^+$ state as a member of the ground-state band. The results for ($\beta_p$, $\beta_n$, $\beta_m$) are (0.185, 0.149, 0.166) for $^{54}$Fe and (0.149, 0.185, 0.166) for $^{54}$Ni. The deformation increase in the rms radius is 0.026 fm for $^{54}$Fe and 0.014 fm for $^{54}$Ni.

\begin{figure}[t]
    \centering
    \includegraphics[width = 0.5\textwidth]{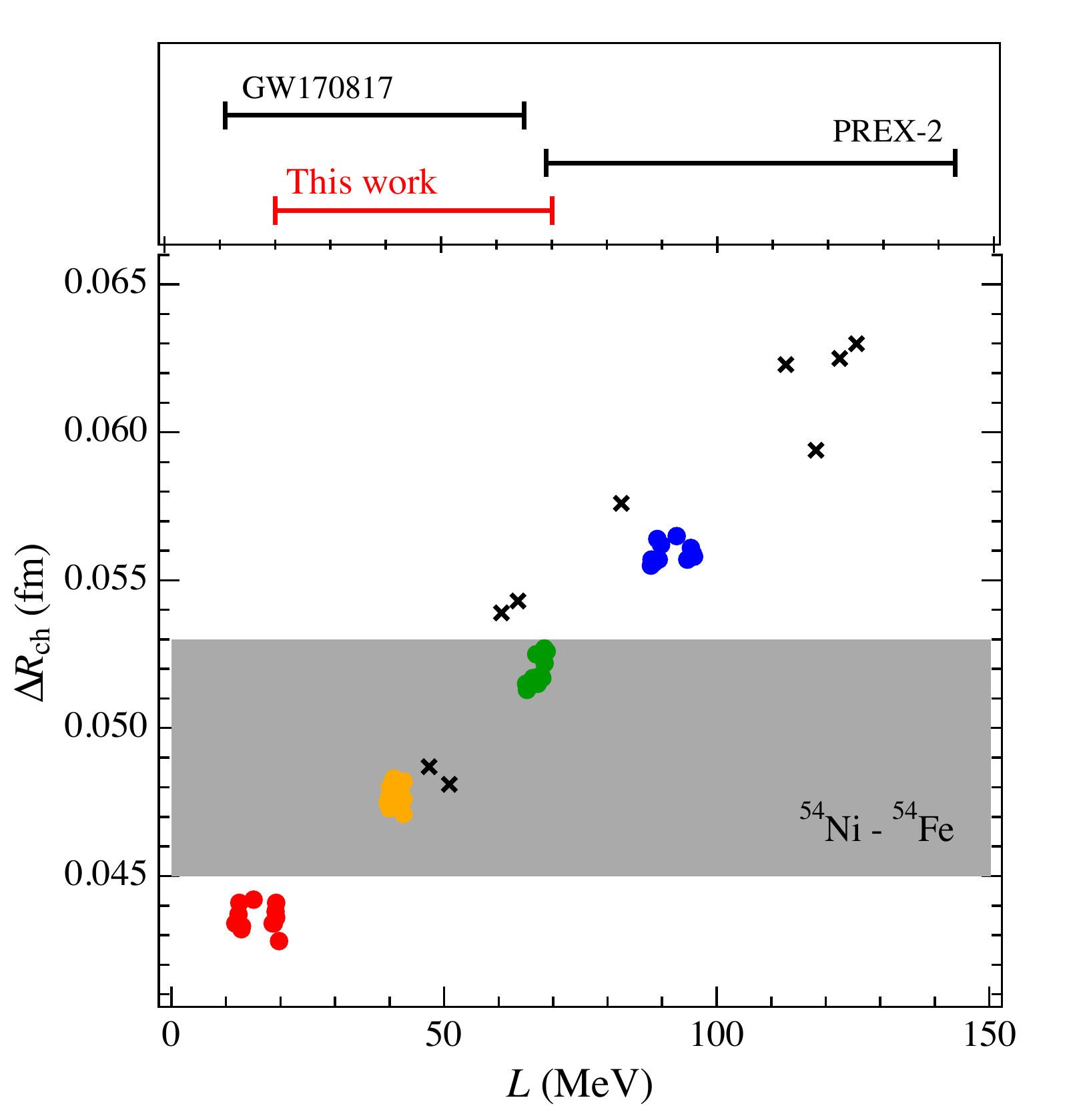}
    \caption{$\Delta R_{\rm ch}$ as a function of $L$ at $\rho_0$. The experimental result is shown as a horizontal gray band. The solid circles are results of Skyrme EDF and the crosses are for the CODF calculations. The upper figure shows comparison with the GW170817 and the PREX-2.}
    \label{fig:LConstraintswoBE2}
\end{figure}

\paragraph{Discussion ---}
The resulting quadrupole correction factor for $\Delta R_{\rm ch}$ is $-0.012$\,fm, which is added to the Skyrme and CODF calculations performed in the spherical basis. If we were to include the higher 2$^+$ states in $^{54}$Ni the quadrupole correction factor would be $-0.010$\,fm. The results are shown in Fig.\,\ref{fig:LConstraintswoBE2} by the colored points. The color indicates the neutron skin of $^{208}$Pb: 0.12\,fm (red), 0.16\,fm (orange), 0.20\,fm (green), and 0.24\,fm (blue) for Skyrme calculations. Also results of the CODF calculations are shown in crosses. The quadrupole correlations are explicitly contained in the CHFB$+$5DCH calculations using the D1S Hamiltonian given in \cite{dec80,ber91}. However, their $  B(E2,\uparrow)_{p}  $ values of 1310 and 1575~e$^{2}$\,fm$^{4}$ for $^{54}$Fe and $^{54}$Ni, respectively, are very different from the $  fp  $ model-space calculations and experiment. 

The Skyrme and CODF calculations show consistent agreement in the correlation between $  \Delta R_{\rm ch}  $ and $  L  $. In comparison to these calculations, the experimental one-sigma error band shown in Fig.\,\ref{fig:LConstraintswoBE2} in gray implies a value of $L$ in the range of 20-70 MeV. In the top panel of Fig.\,\ref{fig:LConstraintswoBE2} we compare the present result with the range for $  L  $ of 11-65 MeV deduced from GW170817\,\cite{rai19}, to which our result is consistent, suggesting a relatively soft neutron matter EOS. The present result is also compared against the recent PREX-2 result of $  \Delta R_{\rm np}  $ = 0.283\,(71) fm\,\cite{adh21} that implies $  L = 106\,(37)  $ MeV\,\cite{Reed:2021nqk}. Our result is barely consistent within 1$\sigma$ error bands with the PREX-2, which indicates rather stiff EOS. It is noted that our previous results on the mirror pair $^{36}$Ca-$^{36}$S indicates the range of $L$ = 5-70 MeV \cite{Brown.2020}, which is consistent with the present results. This implies that the theoretical model dependence is well under control. However, the $A$ = 36 result does not include the quadrupole correlation. It is expected to be small, and once the experimental $B(E2)$ for the $A$ = 36 pair become available, the range from the $A$ = 36 will be updated. 

%The present result is also compared against the recent PREX-2 result of $  \Delta R_{\rm np}  $ = 0.283\,(71) fm\,\cite{adh21} that implies $  L = 106\,(37)  $ MeV\,\cite{Reed:2021nqk}. Our result is consistent with GW170817, suggesting a relatively soft neutron matter EOS, and is also (barely) consistent with the lower limit on $L$ extracted from the PREX-2 measurement. 
\begin{figure}[h]
    \centering
    \includegraphics[width = 0.5\textwidth]{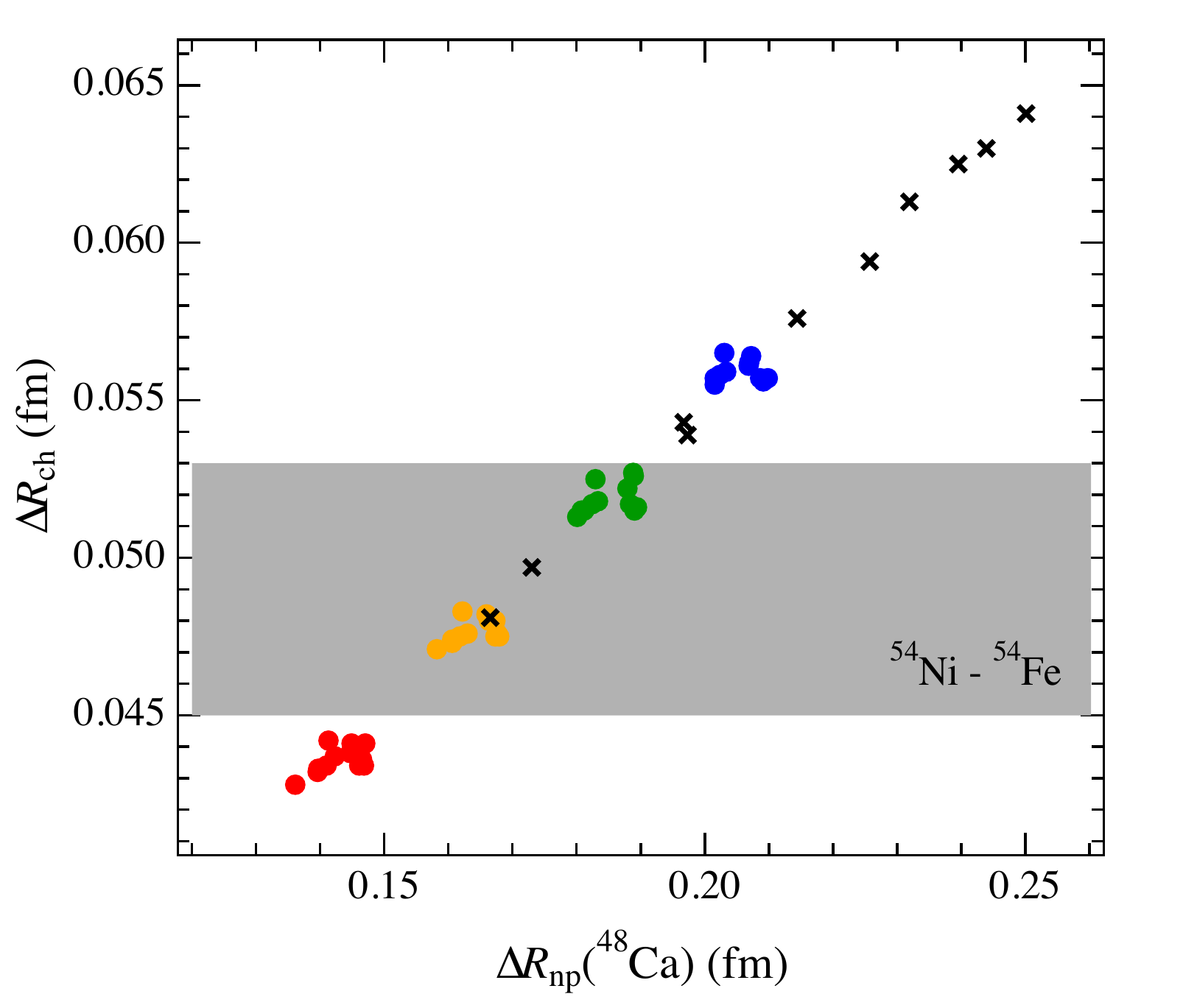}
    \caption{``Data-to-data" relation between $\Delta R_{\rm ch}$ and $\Delta R_{\rm pn}$(${}^{48}$Ca). The same marks and color coding are used as Fig.\,\ref{fig:LConstraintswoBE2}.}
    \label{fig:DatatoData}
\end{figure}

Finally the correlation between $ \Delta R_{\rm ch} $ and $\Delta R_{\rm np}(^{48}{\rm Ca})$ is shown in Fig.\,\ref{fig:DatatoData}. Our $ \Delta R_{\rm ch} $ restricts the  $\Delta R_{\rm np}(^{48}{\rm Ca})$ to the interval of $\!0.15\!-\!0.19\,{\rm fm}$. The connection to ${}^{48}$Ca is timely given that the Calcium Radius EXperiment (CREX) has been completed \cite{CREX:2013}, where experimental error of about $\pm\, 0.02\,{\rm fm}$ is expected, which is comparable to the error obtained here. It is of particular interest whether 
 CREX will confirm the soft EOS or reveal a larger $\Delta R_{\rm np}$ as the PREX-2.

\paragraph{Summary ---}
The $\Delta R_{\rm ch}$ between mirror nuclei $^{54}$Ni-$^{54}$Fe was evaluated, and compared with the Skyrme EDFs and the CODF theories. The $\Delta R_{\rm ch}$ and $L$ correlation implies a range of $L$ = 20-70 MeV, and is consistent with the $L$ from GW170817 and our previous result in the $^{36}$Ca-$^{36}$S pair, suggesting a soft neutron matter EOS. Our result is barely consistent within 1$\sigma$ error bands with the PREX-2 that indicates a stiff EOS. The present $\Delta R_{\rm ch}$ also predicts the $\Delta R_{\rm np}$($^{48}$Ca) as $0.15\!-\!0.19\,{\rm fm}$. More data on the mirror charge radii in different mass regions is required to properly assess the model dependence and to set tighter limits on the $L$. 

\paragraph{Acknowledgements}
This work is support in part by the National Science Foundation grant No. PHY-15-65546 and by the U.S. Department of Energy Office of Science, Office of Nuclear Physics under Award DE-FG02-92ER40750, and by the Deutsche Forschungsgemeinschaft (DFG, German Research Foundation) - Project-ID 279384907 - SFB 1245. We thank Nathalie Pillet for providing the CHFB$+$5DCH calculation results for $^{54}$Ni and $^{54}$Fe.

\bibliography{literature}
\end{document}